# An Experimental Analysis of Work-Life Balance Among The Employees using Machine Learning Classifiers

K. Radha[#1], M. Rohith[#2]

[#1]Assistant Professor, CSE, GITAM (Deemed to be University), Hyderabad, Telangana, India
[#2]B.Tech III-CSE, CSE, GITAM (Deemed to be University), Hyderabad, Telangana, India

**Abstract:**
*Researchers today have found out the importance of Artificial Intelligence, and Machine Learning in our daily lives, as well as they can be used to improve the quality of our lives as well as the cities and nations alike. An example of this is that it is currently speculated that ML can provide ways to relieve workers as it can predict effective working schedules and patterns which increase the efficiency of the workers. Ultimately this is leading to a Work-Life balance for the workers. But how is this possible? It is practically possible with the Machine Learning algorithms to predict, calculate the factors affecting the feelings of the worker's work-life balance. In order to actually do this, a sizable amount of 12,756 people's data has been taken under consideration. Upon analyzing the data and calculating under various factors, we have found out the correlation of various factors and WLB(Work-Life Balance in short). There are some factors that have to be taken into serious consideration as they play a major role in WLB. We have trained 80% of our data with Random Forest Classifier, SVM, and Naïve Bayes algorithms. Upon testing, the algorithms predict the WLB with 71.5% as the best accuracy.*

**Keywords:** *Machine Learning Algorithms, Naïve Bayes, Random Forest Classifier, SVM, Work-Life-Balance.*

## I. INTRODUCTION

What do you think of Artificial Intelligence? Does it scare you to death? Or does it reassure you that there is a brighter future lying ahead of our lives? Well, some sci-fi fans would like to argue that the machines powered by Artificial Intelligence would annihilate us and destroy us all. But more knowledgeable people would like to think the opposite. That Machine Learning is going to empower us with a brighter future. It's forever going to change many areas of our lives for the better. Indeed, the revolutionization of Industry development has advanced the professionals in their professional lives. But still, some would like to argue that mechanization is going to doom us by taking away our jobs, but the changing trend towards concern towards professional's personal lives is a great change in the positive direction.

Recent studies have shown that many employees confessed to switching jobs regarding work-life balance problems. After all, the emergence of Industry 4.0 has led to an increase in a multitude of online communities which are cheap and simple to use, helping employees to work collaboratively on files, documents, conferences, and software. These developments in technology have completely changed the way employees work, causing a change in their work-life balance towards a higher focus on individual lives without jeopardizing work obligations. The globalization of distribution networks has facilitated the need for enhanced collaboration and communication across vast distances and, in many instances, time zones.

Machine Learning may shortly help ease workers of several of their responsibilities, which may enhance efficiency or aid in the discovery of the most efficient methods of performing tasks. As a result, it would help to improve employees' work-life balance in the immediate future. As a direct consequence, workers' standard of living would improve. The potential scenarios could aid in the identification and prevention of work-life imbalance. After analyzing the data from 800 employees in a study, the correlations between certain aspects and work-life balance were discovered. The study was absolutely massive considering the massive population under study and scale of it in comparison to any earlier ones conducted. Overall we can definitely say that work-life balance is generally a balance of one's personal or professional commitments or responsibilities without either's domination above others. But whenever this balance is achieved, one's personal or professional life remains harmonious without any dispute. The demands of work don't preclude any worker from gaining fulfillment from their personal life, and moreover, the personal life and its elements will not overflow and negatively affect their job.

## II. EMPLOYEE ATTRITION

Employee Attrition has a very large interest in today's world. Because it directly effects a particular organization as it loses its good employees due to this. Organizations can incorporate decision-making technologies in a variety of areas: One of the most popularly used innovative technology is Artificial Intelligence that has a wide range of applications like helping clients incorporate strategies, employee management, and organizational aspects. Human Resources has dramatically increased attention in present times. The reason for this might be that businesses have a true competitive edge, and employee's quality and skill embody an increased factor because of HR factors. Artificial Intelligence is now in present times is used to guide workers' management decisions upon being presented to marketing and sales departments within Human Resource management.

By incorporating several machine learning techniques such as Random Forest Classifier, Support Vector Machine(SVM), and even Gaussian Naïve Bayes Classifier and also incorporating several pre-processing methodologies such as feature selection techniques, Exploratory data analysis, also known as EDA, the study's ambition is to identify worker's that are very much likely to quit by taking all of this into account. The construction and enhancement of a great bond between workers and the company can still be assisted while organizations are implementing a range of events to reduce attrition. It's a reality that Employers depending upon their company's priorities, spend a lot of funds and time on worker hiring and training. So this is actually a genuine capital investment for them on the new recruits. So if at any time a worker resigns, they will have resources, HR staff time and also explicitly cash used to invest in them, and most importantly, a very important employee on whom this is all invested. Now we need to know about the meaning of attrition, which generally means that a particular worker from a particular company resigns or retires. But even though this is a procedure, the company should take part in this as the empty available positions should be filled up with new recruits by significantly investing in training and recruitment. This is because though this a very time-consuming and expensive procedure to follow but, this is the only effective way against employee attrition. Some authors advocated that retaining and maintaining strongly driven and genuinely happy workers because they are not only inventive and productive but also enhance the performance, which tends to maintain enhanced organizational value. Furthermore, the backbone of the organization is formed by strongly driven, satisfied, and genuinely satisfied workers, and they also have an effect on the organization's productivity. Strong determinants of displacements and resignations, as seen according to economic literature, is job dissatisfaction and its data, which is also a determinant of turnover intention, also for salaries controlling, standard demographic and factors of job to name a few. In this paper, we have identified the main factors that have an adverse effect on employee attrition and a real classification based on our considered dataset. And we also analyze and understand the reasons, factors, and motivation that drive a worker to leave the company. This data allows the HR department in general to take counter preventive measures such as supplying rewards for employees and an improved working environment. This empowers the HR department by allowing them to embrace management support tools for the workers to use these classification algorithms, which we used here. A real dataset from Gamo Gofa Zone Trade and Industry Development Office from Kaggle is considered for this obtained model for predicting the employee attrition, which includes a sizable amount of 12,756 employees under our study's consideration with 23 factors under consideration initially but later reduced to 16 important factors. A heat map with correlations of 16 factors. With all of these findings, we derive the characteristics that all have very high characteristics for a worker's reason to leave a company. In terms of traditional metrics, we have expressed the results, and the best result of our dataset is given by Support Vector Machine, which gives us the best accuracy in our study by 71.52% accuracy. Overall, this could mean one thing for sure, the information obtained from data analysis clearly shows us that machine learning systems can be adopted for the support of the Human Resources department in the management of company staff.

## III. RELATED WORK

The utility of Human Resource Management(HRM) in operating circumstances, strategic planning, development, and identifying connections with efficiency has been demonstrated by a lot of researchers. As a matter of fact, the effect of HRM on efficiency has a beneficial effect on a company's magnitude, and capital growth effects are confirmed by these findings. Moreover, many of these studies concentrate on analyzing and evaluating customers and their respective behavior and do not discuss a company's main resources, as portrayed by its employees. Many studies have been conducted to investigate employee attrition. An existing study found that workers' trends and attributes that are job-related, such as income and length of employment which play an important role in employee attribute's influence. Another study looked at demographic factors and worker's absenteeism on attrition. The authors concentrated solely on specific job factors. In predicting whether a worker quits or not, the author used a Naïve Bayes Classifier and Decision tree algorithm.

Understanding the factors and impacts of various attributes on the employee's job satisfaction is crucial for a manager to know about and concentrate on so that the future of the organization stays intact. A lot is determined by the management of the organization as this is directly affecting the discontent of salaries or accomplishment, therefore, the involvement of the workers in the improvement of the organization. On multiple events, it is shown that one of the ways to deal with get the best execution out of agents is to make them pleasant and perky. Workers everything being equivalent and levels of pay continue getting dynamically hopeless at work - an extended stretch example that should concern supervisors. How, by then, do directors make satisfied agents? Since pay has been a fundamental thought of pushing delegates in an Organisation. The present investigation broke down the issue of pay and compensation-based satisfaction. Data was accumulated and separated, similar to drawing in experiences, and the analysis of the data was done through the SPSS package program with a 95% confidence interval. A Chi-square test was used to investigate frequencies and explanatory information. The other purpose of this study is to examine the tweets of people who use social media, based on their own declarations, who have a high job satisfaction due to their salary, and those who work with less salary, and as a result of the analysis on tweets sent by these people, the employees who love their job are distinctive. To make clear differences. In this study, data mining, machine learning, and data science methods are applied to data captured on Twitter. A total of 142,656 tweets were made, and these tweets were worked on.

Among the applied algorithms, the highest success rate belongs to the Gradient Boosted Tree, and it can accurately classify on a nearly balanced and bipolar dataset with success over 99%. As a result of the applied machine learning algorithms, it was observed that there was a significant difference between the sentence structures, the words used, and the posts made by those who were satisfied and dissatisfied with the salary they received. Thanks to machine learning, people who have high job satisfaction and who do not have high job satisfaction can be learned by analyzing their Twitter accounts. Other machine learning algorithms can also be attempted when sufficient hardware capacity is available. Techniques have been used to test the importance of association among pay and compensation-based satisfaction. The result demonstrated that there is no significant association between compensation and delegate occupation satisfaction among the respondents. In any case, portions of pay, for instance, livelihood improvement and business dependability, were significant contributing factors to delegate occupation satisfaction. The assessment recommended that better conveyor headway openings should be given to the agents to assemble work satisfaction [6].

### A. Employee Benefits in the Organisations

As indicated by Armstrong (2006), representative advantages incorporate benefits, debilitated compensation, protection spread, organization autos, and various other 'advantages.' Advantages have been utilized to perceive extraordinary commitment, execution, the responsibility to culture, and qualities. Advantages incorporate extra downtime, passes to occasions, trips, meals, open acknowledgment. Also, financial benefits for representatives and representatives that are not carefully compensated, for example, yearly occasions. A few advantages are ordered by law, for instance, government disability, joblessness remuneration, and laborer pay. Representatives' advantages incorporate annuity, medical coverage, incidental advantages, welfare, and others (Lee et al., 2006). Moreover, advantages can treat as the installment or privilege, for example, one makes under a protection approach or work understanding, or open help program or all the more, for the most part, something of significant worth or handiness. Advantages may likewise view as an impression of equity in the public arena (Herman, 2005). As per Carter (2008), Benefits are types of significant worth, other than installment, that are given to representatives as a by-product of their commitment to the association for doing their work. Zhou et al. (2009) contended that Advantages gives adaptable and advertise focused medical advantages to help work brand and bolster fascination and maintenance. Uppal (2005) utilized a measure that involved the quantity of incidental advantages representatives get and finds this has emphatically identified with work fulfillment. Plus, work fulfillment has ascended while benefits fulfillment has raised. The arrangement of these two patterns keeps on recommending that advantages may help laborers content with their employment. Of the individuals who are happy with their advantages, 75% are happy with their present place of employment, contrasted with simply 25% who are happy with their occupations among the individuals who are not happy with their advantages. As managers center on holding representatives, one potential methodology is in the first place remuneration. Strong advantages offering joined with successful advantages training could improve laborers' general advantages fulfillment, which may convert into more prominent occupation fulfillment. All these are ground-breaking pieces of evidence to demonstrate the connection between remuneration and workers' activity fulfillment. Liberal prizes (pay) will, in general, hold individuals since high price levels lead to high occupation fulfillment, duty, and devotion (Chiu et al., 2002). Accordingly, when representatives feel they have not remunerated as they expected, it will diminish their activity fulfillment, inspiration may endure, prompting low confidence and low-quality execution. For instance, if one saw decency about the advantages that one got from one's boss, this may prompt higher occupation fulfillment.

The four components of remuneration, which are base-pay, variable compensation, advantages, and work-life balance are the most well-known offered to workers. Here, let utilize a few instances of parts of remuneration to carry its association with representatives' activity fulfilment. Base-compensation has distinguished as a "cleanliness factor" and all things considered this could cause worker work disappointment if a base pay desire isn't understood (Joseph and Robert, 1995). De Vaney and Chen (2003) recommended that pay or pay as one of the primary variables that altogether impacts representatives' activity fulfilment. Furthermore, there are two components of advantages which are stipends, and work-life balance has all the more frequently offered to representatives. Here, let utilize a few instances of segments of advantages to carrying its association with workers' activity fulfilment. Because of Bonner (1997), health programs including benefits, transport remittance, therapeutic stipend, and so forth, have positive associations with work fulfilment of representatives. After effects of the examination uncovered that accessibility of recompenses significantly affected employment fulfilment. These outcomes are steady with Onu et al. (2005) study in Nigeria who saw that propitious states of administration are significant parts of employment fulfilment. Work-life balance programs have exhibited to affect representatives regarding enrollment, maintenance or turnover, responsibility and occupation fulfilment, non-appearance, efficiency and mishap rates (Melissa, 2007). Also, work fulfilment is seen as contrarily related to work to family impedance. Burke and Glass (1999) found that the representatives have progressively fulfilled and focused on their activity if associations are steady of work-life balance.

## IV. STATEMENT OF THE PROBLEM

In our contempory conservative financial environment, for retaining talented employees and enhancing economic performance, reward systems are becoming an extremely prevelant tool. As a result, the most global organisation understands the role of compensation (Wayne-F- Casio, 2006). Commonly, the employee often seeks different benefits to be more productive and achievable organisational objective. Compensation is any financial and non-financial benefit that can be given to employees as a reward to motivate them to be more productive and recognize their contribution forwards organisations goal achievement and Compensation is a vital tool in helping to encourage employees to be able to work. It is acceptable that in the different organisation as well as Compensation play an important role in increasing satisfaction to employees. The organisation objectives can't be achieved without satisfied employees, but, most organisations in Ethiopia have little understanding of the role of compensation on employee's job satisfaction. As a result, many employees are changing their organisation to get attractive compensation. Among this Gamo Gofa zone, Trade and Industry Development Office in Arba Minch is the one who has lack of awareness on the main roles of compensation on employee's job satisfaction and little understanding on organisational objectives can't achieve without satisfied employees. So, the researcher in this paper focus to investigate compensation which is the best means of creating awareness about the advantage of compensation and employee satisfaction towards their jobs and it also tries to identify the effect of compensation on the employee's job satisfaction and to create awareness about the advantage of compensation in a case of Gamo Gofa zone Trade and Industry Development Office in Arba Minch City.

From the above description of the problem, the following leading questions have raised in the research.

1) How Gamo Gofa Zone Trade and Industry Development office compensate their employees?
2) Are the employees satisfied or not with the existing compensation system?
3) What are the influences of compensation in the organisation?
4) Does the organisation encourage the compensation strategy? General Objective

### A. The Objective of the study

The general objective of the study was to know the role of compensation on employees' job satisfaction the case of Gamo Gofa Zone Trade and Industry Development Office. Specific Objectives

1) To assess how Gamo Gofa zone Trade and Industry Development Office compensate their employees.

2) To find out whether the employees are satisfied or not with existing compensation

3) To examine the impact of compensation in the organisation.

4) To know whether the organisation encourages the compensation strategy

5) To know whether their employees are having a Work-Life-Balance in their lives.

## V. METHODOLOGY STEPS

1) Detection of Data: At this stage, appropriate twitter accounts have been determined to draw data specifically for the problem and twitter accounts have been determined for polar classes (relatively satisfied and dissatisfied with the salary received).
2) Collection of Data: At this stage, data was collected from the accounts determined in the previous stage by providing a connection to Twitter via the Twitter API.
3) Classification of Data: At this stage, data mining techniques were used, feature extraction was made on the data, and then the label of those who were satisfied and dissatisfied with the relative salary received for the data was added.
4) Algorithm Selection: At this stage, tests have been performed on data mining algorithms and adjustments have been made in order to select the appropriate algorithm and to make the selected algorithm produce more successful results.
5) Evaluation of the Results: At this stage, the success of the algorithm results obtained was examined and the evaluation process was continued in a cycle until the satisfactory results were obtained by working on the algorithms according to the values in the results.
Total number of tweets taken in this study analyzed 142,656.

## VI. DETECTION OF DATA

With a bipolar approach, employees, who can be found in two extreme poles, who are satisfied and dissatisfied with the salary they receive, are listed using the internet and their twitter accounts are determined. In determining these accounts, individuals' own statements were taken as basis and at this stage they were made manually by following a researcher's twitter accounts and reading their statements. In the bipolar approach, individuals were not required.

To investigate the causes of employee attrition and based on the phases outlined below, to develop a predictive model (see Figure 1):

1) Firstly, we should collect the dataset of the worker, and this should also consider the dataset that should contain present and past findings.
2) Get the dataset ready by performing data cleaning methods.
3) Now we should do descriptive analysis at this step to find out all the attributes that effect the attrition.
4) Now, we should apply the various classification algorithms and test and train the dataset with them.
5) Now we compare the test results in this case. We compare all the performance metrics this stage of all the classification algorithms which model best fits our requirement and when this is done, and the software that implements the classification model, release it to the HR.

## VII. PRE-PROCESSING

### A. Data Description:

We got our HRM dataset from Kaggle. This dataset initially contains 23 features later narrowed down to 16 with a dataset of 12,756 employees. All of this data is relating to worker's personal and professional characteristics. (see fig 1).

**Table 1.** Dataset features.

```
Data columns (total 23 columns):

Timestamp            12756 non-null object
FRUITS_VEGGIES       12756 non-null int64
DAILY_STRESS         12756 non-null object
PLACES_VISITED       12756 non-null int64
CORE_CIRCLE          12756 non-null int64
SUPPORTING_OTHERS    12756 non-null int64
SOCIAL_NETWORK       12756 non-null int64
ACHIEVEMENT          12756 non-null int64
DONATION             12756 non-null int64
BMI_RANGE            12756 non-null int64
TODO_COMPLETED       12756 non-null int64
FLOW                 12756 non-null int64
DAILY_STEPS          12756 non-null int64
LIVE_VISION          12756 non-null int64
SLEEP_HOURS          12756 non-null int64
LOST_VACATION        12756 non-null int64
DAILY_SHOUTING       12756 non-null int64
SUFFICIENT_INCOME    12756 non-null int64
PERSONAL_AWARDS      12756 non-null int64
TIME_FOR_PASSION     12756 non-null int64
DAILY_MEDITATION     12756 non-null int64
AGE                  12756 non-null object
GENDER               12756 non-null object
```

**Fig: 1 - Dataset columns**

### B. Data Cleaning:

Among the most crucial components of machine learning is data preparation. This process is usually multifactored and takes a vast amount of time.

## VIII. DATA EXPLORATION

### A. Descriptive analysis:

Initial step of descriptive analysis is to look at how the target variable is distributed across the dataset. Our sample has 12,756 employee's data under consideration. And the 16 variables we have take under considerations are- DAILY_STRESS,PLACED_VISITED,SOCIAL_NETWORK, BMI_RANGE, TODO_COMPLETED, DAILY_STEPS, SLEEP_HOURS,LOST_VACATION,DAILY_SHOUTING, SUFFICIENT_INCOME,TIME_FOR_PASSION,DAILY_MEDITATION, AGE, GENDER, GENDER_ENCODER, AGE_ENCODER. We have training 80% of the dataset and testing the remaining 20%.

### B. Data Descriptive Statistics:

dt.head()

DAILY_STRESS

PLACES_VISITED

SOCIAL_NETWORK

BMI_RANGE

TODO_COMPLETED

DAILY_STEPS

SLEEP_HOURS

LOST_VACATION

DAILY_SHOUTING

SUFFICIENT_INCOME

TIME_FOR_PASSION

DAILY_MEDITATION

AGE

GENDER

Gender_Encoder

Age_Encoder

**Fig: 2 - Selected Dataset columns**

```
dt.describe().columns

Index(['FRUITS_VEGGIES', 'PLACES_VISITED', 'CORE_CIRCLE', 'SUPPORTING_OTHERS',
       'SOCIAL_NETWORK', 'ACHIEVEMENT', 'DONATION', 'BMI_RANGE',
       'TODO_COMPLETED', 'FLOW', 'DAILY_STEPS', 'LIVE_VISION', 'SLEEP_HOURS',
       'LOST_VACATION', 'DAILY_SHOUTING', 'SUFFICIENT_INCOME',
       'PERSONAL_AWARDS', 'TIME_FOR_PASSION', 'DAILY_MEDITATION',
       'Gender_Encoder', 'Age_Encoder'],
      dtype='object')

dt.drop(columns=['CORE_CIRCLE','SUPPORTING_OTHERS','ACHIEVEMENT','DONATION','FLOW'
,'LIVE_VISION','PERSONAL_AWARDS','Timestamp','FRUITS_VEGGIES'],inplace=True)
```

**Fig: 3 – Total Dataset columns**

Before moving on to descriptive analysis we should complete categorical encoding after the exploration phase and data cleaning were done. But we need to shift the format of certain variables to encourage better similarity and readability with other evaluated quantities to avoid uncertain results. The format change entailed organising values in clusters in uniform way, to make them all very similar to compare, or visualization every value to a correlating categorical value.

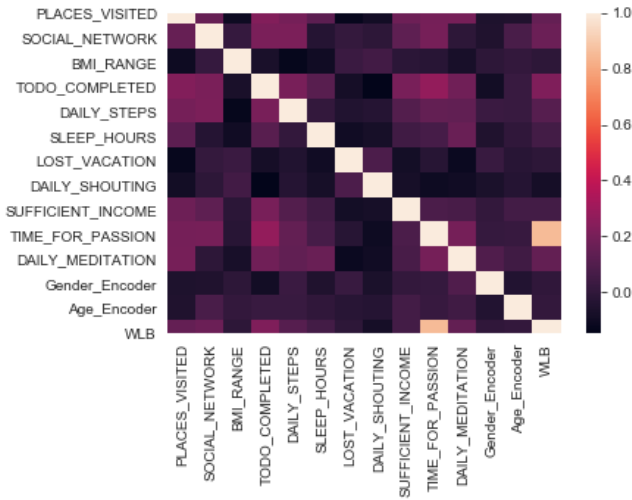

**Fig: 4 - Correlation Heatmap**

In general a heatmap is a graphical representation of data in which the data values are represented with colors. It comes very handy to the users for understanding the data (large data specifically). Correlation heatmap makes a heatmap correlating every column of data w.r.t every other column. Here we made a correlation heatmap of our data by considering every column value against itself and every other column value and the resultant is the heatmap obtained.

## IX. MODEL BUILDING
In this process we basically select machine learning techniques which are suitable in our case. Our primary goal is to identify which predictive model is best suitable for our requirement. We have taken Random Forest Classifier, Support Vector Machine, and Gaussian Naïve Bayes

Classifier into account and each of them should be trained and tested. Our considered algorithms are:

1) Random forest classifier.
2) Support Vector Machine (SVM).
3) Gaussian Naïve Bayes Classifier.

### A. Random Forest Classifier

Classification report:

|  | Precision | Recall | F1-Score | Support |
|---|---|---|---|---|
| 0 | 0.75 | 0.89 | 0.81 | 7325 |
| 1 | 0.45 | 0.22 | 0.3 | 2880 |
| accuracy |  |  | 0.7 | 10205 |
| macro avg | 0.6 | 0.56 | 0.56 | 10205 |
| weighted avg | 0.66 | 0.7 | 0.67 | 10205 |

**Fig:5 – Classification Report Of Random Forest Classifier**

Random Forest Classifier is an ensemble model classification algorithm. Which means that it consists of n decision trees. Also, each of these trees are individual. They give out a class prediction as well. And the majority becomes the final prediction of the entire model. Here we used Random Forest Classifier algorithm which internally is building decision trees based on the data points. For new data points, it is finding the predictions of each decision tree and assigning the new data points to the majority votes category that wins the votes. We have trained 80% data here and based on the result, by taking all considered column values on x values against column- "WLB" on y axis, we got an accuracy score of 70.4%.

### B. Support Vector Machine (SVM) Classifier
Once we identify what our objectives are that is, determining if an employee is having a Work-Life Balance or not, which in turn determines if he/she will leave the company, and after preparing and examining the dataset, we need to go with designing the prediction model. And as we are using supervised learning algorithms in our case, we need to have 2 sets in our dataset- Testing dataset and Training dataset. Training dataset is required to because it already contains classified target population which is in turn needed get new observations. And now that we have it, we need to have a Testing dataset to further refine the prediction capability, and we need to test a fixed and consistent amount against our attributes. And it's a fact that the prediction ability of a machine learning algorithm is drastically increased if we have a large Training dataset to work with. So for this very reason we have internally divided our dataset in 80:20 ratio, where 80% of our dataset is trained and remaining 20% is tested against(See Fig: 7).

Classification report:

| | Precision | Recall | F1-Score | Support |
|---|---|---|---|---|
| 0 | 0.72 | 1 | 0.83 | 7299 |
| 1 | 0 | 0 | 0 | 2906 |
| accuracy | | | 0.72 | 10205 |
| macro avg | 0.36 | 0.5 | 0.42 | 10205 |
| weighted avg | 0.51 | 0.72 | 0.6 | 10205 |

**Fig: 6 – Classification Report Of Support Vector Machine**

Support Vector Machine(SVM) is one of the supervised learning algorithms in Machine Learning. It also does classification and regression analysis of the dataset which is useful for our study. SVM perform linear classification, non-linear classification, and mapping their inputs into high-dimensional feature spaces implicitly. Here we have used linear SVM which is finding the closest points of lines from the two classes X and Y which are called support vectors. The SVM is maximising the distance between the hyperplane and the vectors called as margin. By this calculation, by taking all considered column values on X values against column- "WLB" on y axis, we got an accuracy score of 71.5%.

And now, Train dataset contains 80% of our dataset, and this is important for our prediction model to study an understand the relationships between various attributes in our dataset of 10,205 observations;

1) Test set contained the remaining 20%. Which is important as well to for validating our prediction model, to calculate errors between our predicted and actual results, and in general performance of our prediction model with our remaining 2,551 observations.

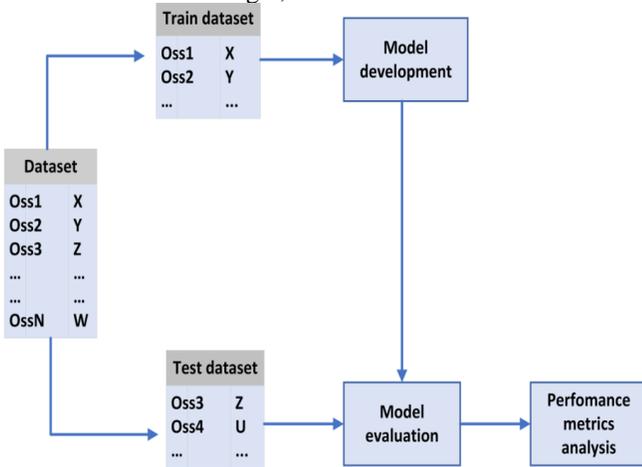

**Fig: 7 - Dataset Split Process**

Now that we have the test and train datasets, to get the target variable attribute(WLB) we can further divide into into 2 sets, X and Y. X dataset contains all the all the variables that are independent and with which we should compare the attrition variable (WLB), and Y dataset contains the WLB (attrition) variable itself. And we need to predict the Y variable through the classification model.

1) X contains the attributes:

DAILY_STRESS,PLACED_VISITED,SOCIAL_N ETWORK, BMI_RANGE, TODO_COMPLETED, DAILY_STEPS, SLEEP_HOURS,LOST_VACATION,DAILY_SH OUTING, SUFFICIENT_INCOME,TIME_FOR_PASSION,D AILY_MEDITATION, AGE, GENDER, GENDER_ENCODER, AGE_ENCODER.

2) Y contains: WLB attribute.

But just this isn't alone sufficient. We need to adapt the Holdout technique as well. This is to better assess and to maintain the target variable to get evenly distributed.
**Holdout:** To maintain the target variable to get evenly distributed when we have the splitted our dataset into test dataset and train dataset. Also this takes care of avoiding disjointing subdivisions which can affect our prediction results. The target (attrition defining attribute- "WLB") is a binary variable with 72% "0" which signifies NO and 28% "1" which signifies YES.

### C. Naïve Bayes
In this classification algorithm, we took advantage of the classification report comprising of Precision, Recall, F1-Score and Support for the algorithm's performance analysis. Also have similarly trained 80% of our dataset and test the remaining 20% to obtain an accuracy score of 70.15%.

Classification report:

| | Precision | Recall | F1-Score | Support |
|---|---|---|---|---|
| 0 | 0.72 | 0.85 | 0.78 | 7299 |
| 1 | 0.29 | 0.16 | 0.2 | 2906 |
| accuracy | | | 0.65 | 10205 |
| macro avg | 0.5 | 0.5 | 0.49 | 10205 |
| weighted avg | 0.59 | 0.65 | 0.61 | 10205 |

**Fig: 8 – Classification Report Of Naïve Bayes Algorithm**

Naïve Bayes Classifiers are based on Bayes Theorem. It is a family of algorithms where all of them share a common principle, i.e. every pair of features being classified is independent of each other. It is based on Naïve Bayes

$$P\left(\frac{B}{A}\right) = \left(\frac{P(A \cap B)}{P(A)}\right)$$

Formula:

Here we are using multinomial Naïve Bayes algorithm, the algorithm's calculation is based on frequency.

By using this Naïve Bayes Algorithm by taking all considered column values on x values against column-"WLB" on y axis, which is calculating the revised probability of a data point to be plotted occurring after taking new information of another axis into consideration.
With this we got an accuracy score of 70.1%.

Cross-validation: It is important to cross validate so that we can be sure that the model is neither over-fitting nor under-fitting.

Here, we divide our dataset into 5 random ones, for testing and training purposes. And the final output error predicted is the mean of all predicted errors of the 5 iterations in this case (Refer Fig: 9).

By this classification algorithm we got an accuracy score of 70.15%.

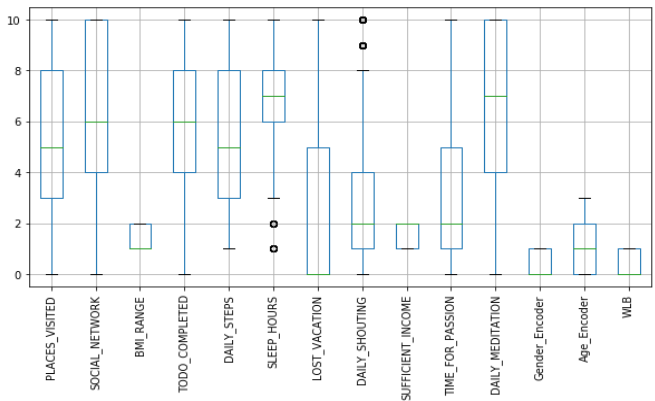

**Fig: 9 - K-Fold Cross-Validation**

## X. DATA VISUALIZATION

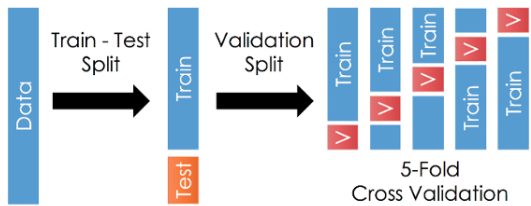

**Fig: 10 – Box Plot**

Here we box plotted the data with all corresponding column values. By this box plot, we got a box from the first quartile to the third quartile corresponding all column values.

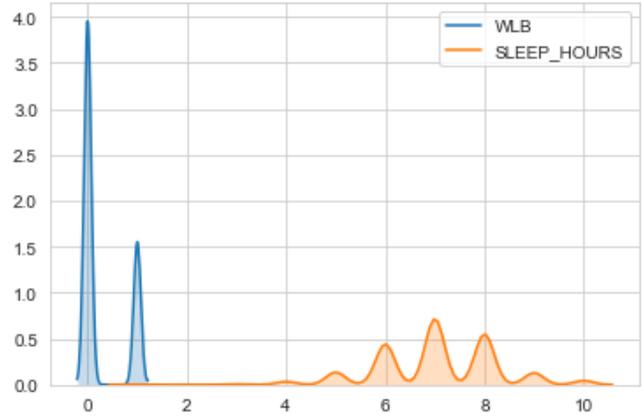

**Fig: 11 - KDE Plot**

By kernel density estimate function on seaborn we plotted a graph with WLB column and Sleep_Hours values against 0.0 to 4.0 graph range. It is doing this by generated the data by binning and counting observations and produces a continuous density estimate of our WLB column and SLEEP_HOURS column

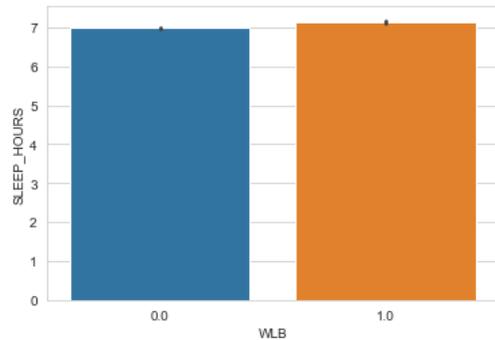

**Fig: 12 - Seaborn Barplot between WLB and SLEEP_HOURS**

With seaborn barplot function we have made point estimates and confidence intervals as rectangular bars thus we made a barplot of WLB column values on x axis w.r.t SLEEP_HOURS values on y axis.

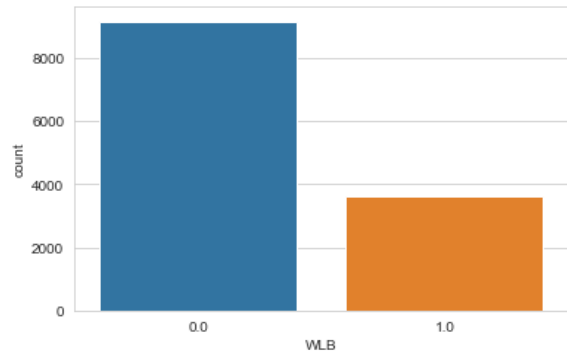

**Fig: 15 – Seaborn Countplot of WLB**

With seaborn countplot function we have made point estimates and confidence intervals as rectangular bars thus we made a countplot of WLB column values on x axis w.r.t count values on y axis.

### A. K means clustering

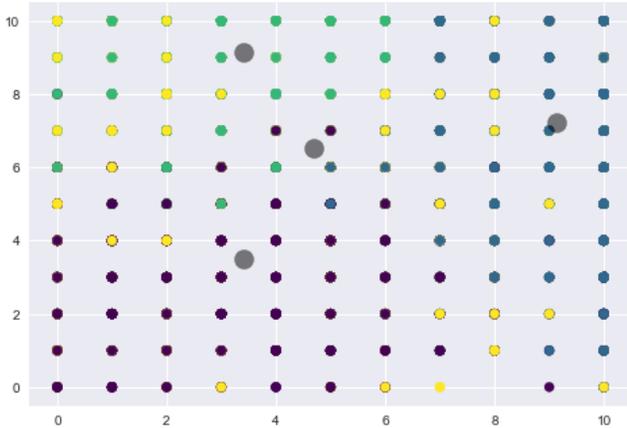

**Fig: 13 – K means clustering**

K Means Clustering is an iterative algorithm and it tries to partition the data into *K* pre-defined distinct non-overlapping clusters where every data point belongs to only one group. *k*-means algorithm searches for a pre-determined no. of clusters within an unlabeled multidimensional dataset.

The *k*-means clustering algorithm is splitting our data set into a 4 clusters and plotting.

With the given data set we plotted a k means cluster above.

### B. Performance Measures

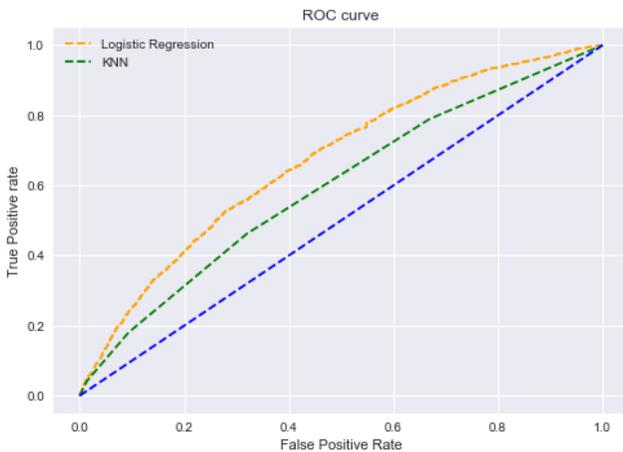

**Fig: 16 - AUC Score and ROC Curve**

Here we plotted a Receiver Operator Characteristic (ROC) curve between Logistic Regression and KNN where ROC curve helps us visualize how well the machine developed is performing.

The ROC curve is showing the trade-off between sensitivity (or TPR) and specificity (1 – FPR). And by comparing the ROC curves, we have taken KNN and Logistic Regression , but the closer curve is of Logistic Regression which only means that among these 2, Logistic Regression has better a better performance.

For satisfying our aim we have successfully applied the machine learning algorithms and derived all the conclusions from them and moreover now are able to predict if an employee will leave the company or not with the help of comparing the WLB factor( If its value is 0 he/she will leave the job, else if it is 1, he/she won't). To get to this we have done a lot of steps, like first statistically assessing the data and later on classifying them. Then, we internally divided our existing dataset into a ratio of 80:20, 80% is trained and 20% is tested. And to maintain the target variable to get evenly distributed, we used the holdout technique. And then, we selected 3 classification algorithms- Random Forest Classifier, Support Vector Machine and Gaussian Naïve Bayes Classifier, and we tested and trained data with them respectively. And we used confusion matrices as well for giving it the predicted results as input and furthermore evaluating the algorithm's performance. Here on, we took advantage of Precision, Recall, Accuracy, F1 Score, AUC curve and ROC curve which are all the basic calculating metrics. Support Vector Machine classifier is the best algorithm in our study as it resulted in the best accuracy score of 71.5% and a recall rate of 1%. From all of this, we are able to derive that SUFFICIENT_INCOME played a major role in determining whether the employee will leave the job or not. And following this, BMI_RANGE, SLEEP_HOURS,DAILY_STRESS,DAILY_MEDITATION, DAILY_SHOUTING too played a major role in determining the same in decreasing priority order. This is definitely the triggering point in the area of predicting whether an employee will leave the job or not, a very important factor for companies and their respective HR departments seriously consider and study. Therefore, proving this study a very interesting and important one for them.

This preliminary study sought to analyze compensation as a source of employee job satisfaction. This examination has endeavoured to research the job of pay on representative's employment fulfilment in Gamo Gofa zone Trade and Industry Development Office. The analyst needs to put the finish of this specific investigation forward to the peruses as follows.

- Concerning this examination, the remuneration procedure urges representatives to try sincerely, and it has its bit of leeway and weakness to the workers just as the association.
- According to, the representative's reaction reasonable and

evenhanded remunerations can give workers work fulfilment, increment association efficiency and abatement divert over from the association. For the most part, reasonable and fair remuneration framework assumes a key job in the association.

## XI. CONCLUSION

In terms of job satisfaction of the employees, the harmonisation of the managers' behaviour, the communication between the subordinate and the superior, and the development of positive working environments will further increase the performance of the employees employed in the organisations and thus can directly affect the increase in the overall business performance.